\documentstyle[aps,prl,twocolumn,epsf]{revtex}
\newcommand{\sizen}{7.5truecm}

\newcommand{\be}{\begin{equation}}
\newcommand{\ee}{\end{equation}}
\newcommand{\ba}{\begin{eqnarray}}
\newcommand{\ea}{\end{eqnarray}}

\newcommand{\Msol}{\mbox{${\rm M}_{\odot}\;$}}

\newcommand{\Msun}{\mbox{${\rm M}_{\odot}\;$}}

\def\lsim{\mathrel{\rlap{
\lower3pt\hbox{\hskip-3pt$\sim$}}
\raise1pt\hbox{$<$}}}

\begin{document}
\preprint{ DRAFT } \title{\bf Prospects of Detecting Baryon and Quark
Superfluidity from Cooling Neutron Stars}

\author {Dany Page$^1$, Madappa Prakash$^2$, James M. Lattimer$^2$,
         and Andrew Steiner$^2$}
\address{
$^1$Instituto de Astronom\'{\i}a, UNAM, Mexico D.F. 04510, Mexico \\
$^2$Department of Physics \& Astronomy, SUNY at Stony Brook,
Stony Brook, NY 11794-3800, USA \\
}
\date{\today}

\maketitle

\begin{abstract}

Baryon and quark superfluidity in the cooling of neutron
stars are investigated.  Observations could constrain
combinations of the neutron or $\Lambda-$hyperon pairing 
gaps and the star's mass.
However, in a hybrid star with a mixed phase of hadrons and quarks,
quark gaps larger than a few tenths of an MeV render quark matter
virtually invisible for cooling.  If the quark gap is smaller, quark
superfluidity could be important, but its effects will be nearly
impossible to distinguish from those of other baryonic
constituents. \\

\noindent PACS number(s): 97.60.Jd. 21.65.+f, 95.30.q \\
\end{abstract}

Pairing is unavoidable in a degenerate Fermi liquid if there is an
attractive interaction in {\it any} channel.  The resulting
superfluidity, and in the case of charged particles,
superconductivity, in neutron star interiors has a major effect on the
star's thermal evolution through suppressions of neutrino
($\nu$) emission processes and specific heats \cite{PA92,P98a}.  Neutron
($n$), proton ($p$) and $\Lambda$-hyperon superfluidity in the $^1S_0$
channel and $n$ superfluidity in the $^3P_2$ channel have been shown
to occur with gaps of a few MeV or less \cite{BEEHJS98,BB97}.
However, the density ranges in which gaps occur remain
uncertain.  At large baryon densities for which perturbative QCD
applies, pairing gaps for like quarks have been estimated to be a few
MeV \cite{BL84}.  However, the pairing gaps of unlike quarks ($ud,~
us$, and $ds$) have been suggested to be several tens to
hundreds of MeV through non-perturbative studies \cite{qsf0}
kindling interest in quark superfluidity and superconductivity
\cite{qsf,B99} and their effects on
neutron stars.

The cooling of a young (age $<10^5$ yr) neutron star is mainly
governed by $\nu-$emission processes and the specific heat
\cite{P98a}.  Due to the extremely high thermal conductivity of
electrons, a neutron star becomes nearly isothermal within a time
$t_w\approx1-100$ years after its birth, depending upon the thickness
of the crust~\cite{LvRPP94}.  After this time its thermal evolution is
controlled by energy balance: 
\be \frac{dE_{th}}{dt} = C_V \frac{dT}{dt} = -L_{\gamma} -L_{\nu} + H \,,
\label{equ:balance}
\ee where $E_{th}$ is the total thermal energy and $C_V$ is the
specific heat.  $L_{\gamma}$ and $L_{\nu}$ are the total luminosities
of photons from the hot surface and $\nu$s from the interior,
respectively.  Possible internal heating sources, due, for example, to
the decay of the magnetic field or friction from differential
rotation, are included in $H$.  Our cooling simulations were performed
by solving the heat transport and hydrostatic equations including
general relativistic effects (see \cite{P98a}).  The surface's
effective temperature $T_e$ is much lower than the internal
temperature $T$ because of a strong temperature gradient in the
envelope.  Above the envelope lies the atmosphere where the emerging
flux is shaped into the observed spectrum from which $T_e$ can be
deduced.  As a rule of thumb $T_e/10^6$ K
$\approx\sqrt{T/10^8{\rm~K}}$, but modifications due to magnetic
fields and chemical composition may occur.

The simplest possible $\nu$ emitting processes are the direct Urca
processes $f_1 + \ell \rightarrow f_2 + \nu_\ell\,, f_2 \rightarrow
f_1 + \ell + \overline{\nu_\ell}$, where $f_1$ and $f_2$ are either
baryons or quarks and $\ell$ is either an electron or a muon.  These
processes can occur whenever momentum conservation is
satisfied among $f_1, f_2$ and $\ell$ (within minutes of
birth, the $\nu$ chemical potential vanishes).
{If the unsuppressed 
direct Urca process for {\em any} component occurs, a neutron star
will rapidly cool because of enhanced emission:
the star's interior temperature $T$ will drop below 10$^9$ K in minutes
and reach 10$^7$ K in about a hundred years}.
$T_e$ will hence drop to less than 300,000 K after
the crustal diffusion time $t_w$ {\cite{PA92,LvRPP94,note}}.  This is the
so-called {\rm rapid cooling} paradigm.  If no direct Urca
processes are allowed, or they are all suppressed, cooling instead
proceeds through the significantly less rapid modified Urca process in
which an additional fermion enables momentum conservation.  This
situation could occur if no hyperons are present, or the nuclear
symmetry energy has a weak density dependence~\cite{LPPH91,PPLP92}.
The $\nu$ emisssion rates for the nucleon, hyperon, and quark Urca
and modified Urca processes can be found in~\cite{crevs}.

The effect of the pairing gaps on the emissivities and specific heats
for massive baryons are investigated in \cite{LY9496} and are here
generalized to the case of quarks.  The principal effects are severe
suppressions of both the emissivity and specific heat when
$T<<\Delta$, where $\Delta$ is the pairing gap.  In a system in which
several superfluid species exist the most relevant gap for these
suppressions is the smallest one.  The specific heat suppression is
never complete, however, because leptons remain unpaired.  Below the
critical temperature $T_c$, pairs may recombine, resulting in the
emission of $\nu\bar\nu$ pairs with a rate that exceeds the modified
Urca rate below $10^{10}$ K~\cite{FRS}; these processes are included
in our calculations.

\begin{center}
  \begin{figure} \epsfxsize \sizen
     \epsfbox{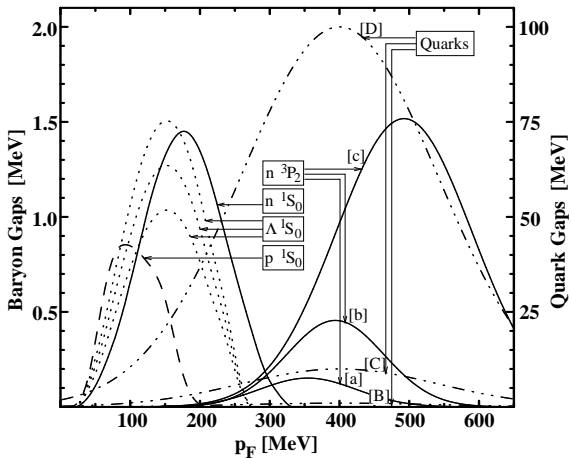}
     \caption{Pairing gaps adopted for neutron $^1$S$_0$ and $^3$P$_2$,
     proton $^1$S$_0$, $\Lambda$ $^1$S$_0$, and quarks.
     The n $^3$P$_2$ gaps are anisotropic; plotted values are
     angle-averaged.    The $\Lambda$ gaps
correspond, in order of increasing $\Delta$, to background densities
$n_B=0.48, 0.64$ and 0.8 fm$^{-3}$, respectively. 
     The s-wave quark gaps are schematic; see text for details.}
\label{fig:baryon_Tc} \end{figure}
\end{center}

The baryon and quark pairing gaps 
we adopt are shown in
Fig.~\ref{fig:baryon_Tc}.  Note that gaps are functions of Fermi
momenta ($p_F(i)$, $i$ denoting the species) 
which translates into a density dependence.  For $p_F(n,p)
\lsim 200 - 300$ MeV/$c$, nucleons pair in the $^1$S$_0$ state, but
these momenta correspond to densities too low for enhanced $\nu$
emission involving nucleons to occur. At higher $p_F$'s, baryons pair in
higher partial waves.  The $n~^3$P$_2$ gap has been calculated for the
Argonne V$_{18}$, CD-Bonn and Nijmegen I \& II interactions
\cite{BEEHJS98}.  This gap is crucial since it extends to large $p_F(n)$
and can reasonably be expected to occur at the centers of neutron
stars.  For $p_F(n)>350$ MeV/c, gaps are largely uncertain because of
both experimental and theoretical uncertainties \cite{BEEHJS98}.  The
curves [a], [b] and [c] in Fig.~\ref{fig:baryon_Tc} reflect the range
of uncertainty.  The $p$ $^3$P$_2$  gap is too small to be
of interest.  Gaps for the $^1$S$_0$ pairing of $\Lambda$, taken from
\cite{BB97} and shown as dotted curves, are highly relevant since
$\Lambda$s participate in direct Urca emission as soon as they appear
\cite{PPLP92}.  Experimental information beyond the $^1$S$_0$ channel
for $\Lambda$ is not available.  $\Delta$s for $\Sigma-$hyperons
remain largely unexplored. The quark ($q$) gaps are taken to be Gaussians
centered at $p_F(q) = 400$ MeV/$c$ with widths of 200 MeV/$c$ and heights
of 100 MeV [model D], 10 MeV [C], 1 MeV [B] and 0.1 MeV [A],
respectively.  The reason for considering quark gaps much smaller than
suggested in \cite{BL84,qsf0} is associated with the
multicomponent nature of charge-neutral, beta-equilibrated, neutron
star matter as will become clear shortly.

We consider four generic compositions: charge-neutral, beta
equilibrated matter containing nucleons only ($np$), nucleons with
quark matter ($npQ$), nucleons and hyperons ($npH$), and nucleons,
hyperons and quarks ($npHQ$).  In the cases involving quarks, a mixed phase
of baryons and quarks is constructed by satisfying Gibbs' phase rules
for mechanical, chemical and thermal equilibrium \cite{G92}.  The
phase of pure quark matter exists only for very large baryon
densities, and rarely occurs in our neutron star models.  Baryonic
matter is calculated using a field-theoretic model at the mean field
level \cite{ZM90}; quark matter is calculated using either a bag-like
model or the Nambu-Jona-Lasinio quark model \cite {PCLSPL}. 
The
equation of state (EOS) is little affected by the pairing phenomenon,
since the energy density gained 
is negligible compared to the ground state energy densities without pairing.

Additional particles, such as quarks or hyperons, have the
effect of softening the EOS and increasing the central densities of
stars relative to the $np$ case.  
For the $npQ$ model studied, a mixed phase appears at
the density $n_B=0.48$ fm$^{-3}$.  Although the volume fraction of
quarks is initially zero, the quarks themselves have a significant
$p_F(q)$ when the phase appears.  The $p_F$s of the three quark
flavors become the same at extremely high density, but for the densities
of interest they are different due to
the presence of negatively charged leptons. In particular, $p_F(s)$
is much smaller than $p_F(u)$ and $p_F(d)$ due to the
larger $s$-quark mass.  Use of the Nambu--Jona-Lasinio model, in which quarks
acquire density-dependent masses resembling those of constituent
quarks, exaggerates the reduction of $p_F(s)$.  This has
dramatic consequences since the pairing phenomenon operates at its
maximum strength when the Fermi momenta are exactly equal; even small
asymmetries cause pairing gaps to be severely reduced
\cite{B99,ARSW96}.  In addition, one may also expect p-wave
superfluidity, to date unexplored, which may yield gaps smaller than
that for the s-wave.  We therefore investigate pairing gaps that are
much smaller than those reported for the case of s-wave superfluidity
and equal quark $p_F$'s.  

The introduction of hyperons does not change these generic trends.  In
the case $npH$, the appearance of hyperons changes the lepton and
nucleon $p_F$'s similarly to the appearance of quarks 
although with less magnitude.  
While the appearance of quarks is delayed by the
existence of hyperons, at high densities the $p_F$'s of nucleons and
quarks remain similar to those of the $npQ$ case.  The existence of either
hyperons or quarks, however, does allow the possibility of additional
direct Urca processes involving themselves as well as those involving
nucleons by decreasing $p_F(n)-p_F(p)$.  For the $npQ$ and $npHQ$
models studied, the maximum masses are $\cong 1.5$\Msol, the central
baryon densities are $\cong 1.35$ fm$^{-3}$, and the volume fractions
of quarks at the center are $\cong 0.4$.

\begin{center}
  \begin{figure} \epsfxsize \sizen \epsfbox{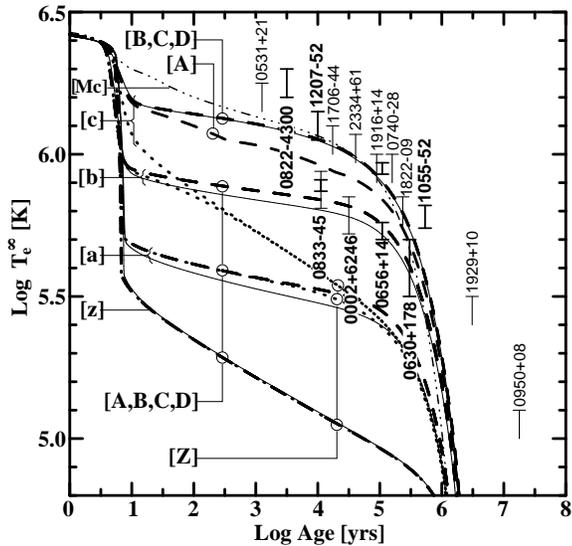}
     \vspace{0.1in} \caption{Cooling of 1.4\Msun stars with $np$
     matter (continuous curves) and npQ matter (dashed and dotted
     curves).  The curves labelled as [a], [b], and [c] correspond to
     $n~^3$P$_2$ gaps as in Fig.~\protect\ref{fig:baryon_Tc}; [z]
     corresponds to zero $n$ gap.  Models labelled [A], [B], [C] and
     [D] correspond to quark gaps as in
     Fig.~\protect\ref{fig:baryon_Tc}; [Z] corresponds to zero quark
     gap.} \label{fig:cooling-N-NQ} \end{figure}
\end{center}

\begin{center}
  \begin{figure} \epsfxsize \sizen
     \epsfbox{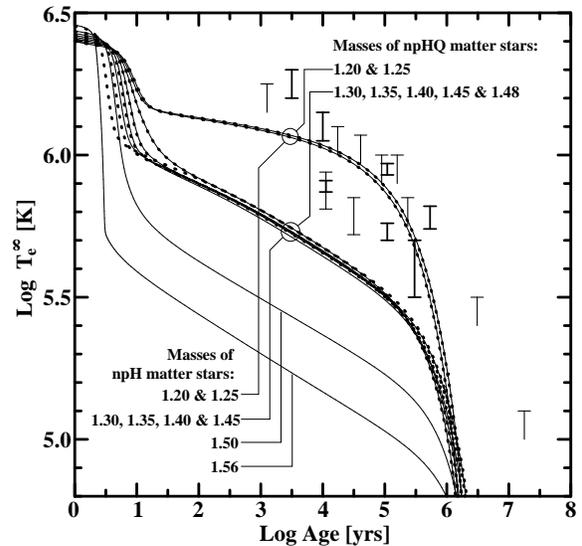} \caption{Cooling of stars
     with $npH$ (continuous lines) and $npHQ$ matter (dotted lines)
     for various stellar masses (in \Msun).  $n~^3$P$_2$ gaps are from
     case [c] while quark gaps, when present, are from model [C] of
     Fig.~\ref{fig:baryon_Tc}.} \label{fig:cooling-NH-NHQ}
     \end{figure}
\end{center}

Cooling simulations of stars without hyperons and with hyperons are
compared, in Figs.~\ref{fig:cooling-N-NQ} and
\ref{fig:cooling-NH-NHQ}, respectively, to available observations of
thermal emissions from isolated neutron stars.  
Sources for the observational data can be found in \cite{P97}.  
However, at the present time, the
inferred temperatures must be considered as upper limits because the
total flux is contaminated, and in some cases dominated, by the
pulsar's magnetospheric emission and/or the emission of a surrounding
synchrotron nebula.  Furthermore, the neutron star surface may be
reheated by magnetospheric high energy photons and particles;
late-time accretion for non-pulsing neutron stars is also
possible.  Other uncertainties arise in the temperature estimates due
to the unknown chemical composition and magnetic field strength in the
surface layers, and in the age, which is based upon the observed
spin-down time.  In these figures, the bolder the data symbol the
better the data.

The $np$ case is considered in Fig.~\ref{fig:cooling-N-NQ}, in which
solid lines indicate the temperature evolution of a 1.4 M$_\odot$ star
for quarkless matter: case [z] is for no nucleon pairing at all, and
cases [a], [b] and [c] correspond to increasing values for the neutron
$^3P_2$ gap, according to Fig.~\ref{fig:baryon_Tc}.  The
field-theoretical model employed for the nucleon interactions allows
the direct nucleon Urca process, which dominates the cooling.  The
unimpeded direct Urca process carries the temperature to values well
below the inferred data.  Pairing suppresses the cooling for $T<T_c$,
where $T$ is the interior temperature, so $T_e$ increases with
increasing $\Delta$.  If the direct Urca process is not allowed, the
range of predicted temperatures is relatively narrow due to the low
emissivity of the modified Urca process. We show an example of such
cooling (curve [Mc]) using the $n~^3P_2$ gap [c] for a 1.4\Msun with
an EOS \cite{APR98} for which the direct Urca cooling is not allowed.

The other curves in the figure illustrate the effects of quarks upon
the cooling.  The dotted curves [Z] are for vanishingly small quark
gaps; the dashed curves ([A], [B], [C] and [D]) are for quark gaps as
proposed in Fig.~\ref{fig:baryon_Tc}.  For nonexistent ([z]) or small
([a]) nucleon gaps, the quark Urca process is irrelevant and the
dependence on the existence or the size of the quark gaps is very
small.  However, for large nucleon gaps ([b] and [c]), the quark
direct Urca process quickly dominates the cooling as the nucleon
direct Urca process is quenched.  It is clear that for quark gaps of
order 1 MeV or greater ([B], [C] or [D]) the effect of quarks is
again very small.  There is at most a slight increase in the stars
temperatures at ages between 10$^1$ to 10$^{5 - 6}$ years due to the
reduction of $p_F(n)$ and the consequent slightly larger gap
(Fig.~\ref{fig:baryon_Tc}).  Even if the quark gap is quite small
([A]), quarks have an effect only if the nucleon gap is very large
([b] or [c]), i.e., significantly larger than the quark gap: the
nucleon direct Urca process is suppressed at high temperatures and the
quark direct Urca process has a chance to contribute to the cooling.
We find that the effects of changing the stellar mass $M$ are similar
to those produced by varying the baryon gap, so that only combinations
of $M$ and $\Delta$ might be constrained by observation.

The thermal evolution of stars containing hyperons has been discussed
in \cite{P98b,SBSB98}, but we obtain qualitatively different results
here.  Hyperons open new direct Urca channels:
$\Lambda \rightarrow p + e + \overline{\nu}_e$ and
$\Lambda + e \rightarrow \Sigma^- + \nu_e$ if $\Sigma^-$'s are present,
with their inverse processes.
Previous results showed that the cooling is
naturally controlled by the smaller of the $\Lambda$ or $n$ gap.
However, this is significantly modified when the
$\Lambda$ gap is more accurately treated.  At the $\Lambda$ appearance
threshold, the gap must vanish since $p_F(\Lambda)$ is vanishingly
small.  We find that a very thin layer, only a few meters thick, of
unpaired or weakly paired $\Lambda$'s is sufficient to control the
cooling.  This effect was overlooked in previous works perhaps because
they lacked adequate zonal resolution.

In Fig.~\ref{fig:cooling-NH-NHQ} we compare the evolution of stars of
different masses made of either $npH$ or $npHQ$ matter.  We find that
all stars, except the most massive $npH$ ones, follow two distinctive
trajectories depending on whether their central density is below or
above the threshold for $\Lambda$ appearance (= 0.54 fm$^{-3}$ in our
model EOS, the threshold star mass being 1.28\Msun).  In the case of
$npH$ matter, stars with $M>1.50\Msun$ are dense enough so that the
$\Lambda$ $^1$S$_0$ gap vanishes and hence undergo fast cooling, while
stars made of $npHQ$ matter do not attain such high densities.  The
temperatures of $npH$ stars with masses between 1.3 and 1.5 \Msun are
below the ones obtained in the models of Fig.~\ref{fig:cooling-N-NQ}
with the same $n$ $^3$P$_2$ gap [b], which confirms that the cooling
is dominated by the very thin layer of unpaired $\Lambda$'s (the
slopes of these cooling curves are typical of direct Urca processes).
Only if the $n~^3$P$_2$ gap $\lsim0.3$ MeV do the cooling curves fall
below what is shown in Fig.~\ref{fig:cooling-NH-NHQ}.  Notice,
moreover, that in the mass range 1.3 -- 1.48 \Msun the cooling curves
are practically indistinguishible from those with unpaired quark
matter shown in Fig.~\ref{fig:cooling-N-NQ}.  In these models with
$npH$ or $npHQ$ matter, there is almost no freedom to ``fine-tune''
the size of the gaps to attain a given $T_e$: stars with $\Lambda$'s
will all follow the same cooling trajectory, determined by the
existence of a layer of unpaired or weakly paired $\Lambda$'s, as
long as the $n~^3P_2$ gap is not smaller.  It is, in some sense, the
same result as in the case of $np$ and $npQ$ matter: the smallest gap
controls the cooling and now the control depends on how fast the
$\Lambda$ $^1$S$_0$ gap increases with increasing $p_F(\Lambda)$.

Our results indicate that observations could constrain combinations of
the smaller of the neutron and $\Lambda-$hyperon pairing gaps and the
star's mass.  Deducing the sizes of quark gaps from observations of
neutron star cooling will be extremely difficult.  Large quark gaps
render quark matter practically invisible, while vanishing quark gaps
lead to cooling behaviors which are nearly indistinguishable from
those of stars containing nucleons or hyperons.  Moreover, it also
appears that cooling observations by themselves will not provide
definitive evidence for the existence of quark matter itself.


Research support (for DP) from grants
Conacyt (\#27987-E) and UNAM-DGAPA (\#IN-119998), from
NSF grant INT-9802680 (for MP and JML) and
DOE grants FG02-88ER-40388 (for MP and AS) and FG02-87ER40317 (for JML) is
gratefully acknowledged.



\begin{thebibliography}{10}
\bibitem{note} The relation between $T_e$ and $T$ is increased by
about 50\% if H or He dominates the atmosphere's composition.  Our qualitative
results concerning superfluidity, however, will be unaffected by the
atmospheric composition.
\end{thebibliography}
\end{document}